\documentclass[pra, onecolumn, longbibliography, floatfix, superscriptaddress]{revtex4-1}
\usepackage{dcolumn}
\usepackage{bm}
\usepackage{graphicx}
\usepackage{amsmath}
\usepackage{enumitem}
\usepackage{latexsym}
\usepackage{amsfonts}
\usepackage{amssymb}
\usepackage{array}
\usepackage{epsfig}
\usepackage[dvipsnames]{xcolor}

\usepackage{color}
\usepackage[colorlinks=true,bookmarks=false,linkcolor=RoyalBlue,urlcolor=RoyalBlue,citecolor=RoyalBlue,breaklinks]{hyperref}
\usepackage{hyperref}
\usepackage{physics}
\usepackage{bbold}

\begin{document}

\newcommand{\mrcomment}[1]{{\textsf{\color{blue} #1}}}

\title{Stochastic collision model approach to transport phenomena in quantum networks}

\author{Dario A. Chisholm}
\affiliation{Universit$\grave{a}$  degli Studi di Palermo, Dipartimento di Fisica e Chimica - Emilio Segr\`e, via Archirafi 36, I-90123 Palermo, Italy}
\affiliation{QTF Centre of Excellence, Turku Centre for Quantum Physics,
Department of Physics and Astronomy, University of Turku, FI-20014 Turun Yliopisto, Finland}
\author{Guillermo Garc\'{i}a-P\'{e}rez}
\affiliation{QTF Centre of Excellence, Turku Centre for Quantum Physics,
Department of Physics and Astronomy, University of Turku, FI-20014 Turun Yliopisto, Finland}
\affiliation{Complex Systems Research Group, Department of Mathematics and Statistics,
University of Turku, FI-20014 Turun Yliopisto, Finland}
\author{Matteo A. C. Rossi}
\affiliation{QTF Centre of Excellence, Turku Centre for Quantum Physics,
Department of Physics and Astronomy, University of Turku, FI-20014 Turun Yliopisto, Finland}
\author{G. Massimo Palma}
\affiliation{Universit$\grave{a}$  degli Studi di Palermo, Dipartimento di Fisica e Chimica - Emilio Segr\`e, via Archirafi 36, I-90123 Palermo, Italy}
\affiliation{NEST, Istituto Nanoscienze-CNR, Piazza S. Silvestro 12, 56127 Pisa, Italy}
\author{Sabrina Maniscalco}
\affiliation{QTF Centre of Excellence, Turku Centre for Quantum Physics,
Department of Physics and Astronomy, University of Turku, FI-20014 Turun Yliopisto, Finland}
\affiliation{QTF Centre of Excellence, Center for Quantum Engineering,
Department of Applied Physics, Aalto University School of Science, FIN-00076 Aalto, Finland}

\begin{abstract}
Noise-assisted transport phenomena highlight the nontrivial interplay between environmental effects and quantum coherence in achieving maximal efficiency. Due to the complexity of biochemical systems and their environments, effective open quantum system models capable of providing physical insights on the presence and role of quantum effects are highly needed. In this paper, we introduce a new approach that combines an effective quantum microscopic description with a classical stochastic one. Our stochastic collision model describes both Markovian and non-Markovian dynamics without relying on the weak coupling assumption. We investigate the consequences of spatial and temporal heterogeneity of noise on transport efficiency in a fully connected graph and in the Fenna-Matthews-Olson complex. Our approach shows how to meaningfully formulate questions, and provide answers, on important open issues such as the properties of optimal noise and the emergence of the network structure as a result of an evolutionary process. 
\end{abstract}
\date{\today}
\maketitle

\section{Introduction}

What is the realm of validity of quantum physics? 
Conceived as the most fundamental physical theory, describing the behavior of the constituents of our Universe, quantum physics has, since its very birth, pushed forward very intriguing fundamental questions of both philosophical and scientific nature. Despite its markedly different predictions with respect to the classical description of reality, speculations on its possible role in key biological processes date back to its founding fathers \cite{Schrodinger2009}. More recently, with the formidable advances in experimental and numerical approaches, fields like quantum biology and quantum complex science have highlighted the existence and persistence of quantum coherence even in complex macroscopic systems. Nonetheless, whether quantumness plays a functional role in biological complexes remains to date a most fascinating open question.

Initial investigations on the quantum measurement problem suggested that the emergence of a classical description of reality from the underlying quantum one could be explained in the framework of environment-induced decoherence \cite{Zurek1991,Zurek2003}. Within this approach, the larger the quantum system, the faster the loss of quantumness due to the interaction with the environment. More recently, however, it has become evident that environmental noise need not be an enemy to the  preservation of quantumness \cite{Verstraete2009}. On the contrary, it may sustain the persistence of quantum coherence and, as a consequence, improve the efficiency of quantum transport in complex systems \cite{Rebentrost2009,Caruso2009,Chin_2010,Uchiyama2017,Rossi2017,Kurt2020}. In this sense, the initial skepticism on the presence of quantum phenomena in macroscopic ``hot and dirty'' systems, due to their very short coherence time, has been overcome \cite{Cao2020}. Nonetheless, complex quantum systems such as, e.g., the extensively studied photosynthetic complexes, are undoubtedly strongly interacting with complex environments, the characteristics of which are very hard to model microscopically.

The enormous challenge of a fully quantum microscopic description of the environment of biological complexes stems both from experimental difficulties in extracting its detailed features and from the exponential increase in the resources needed to simulate quantum many-body systems of large size. Moreover, standard approaches of open quantum systems theory rely on approximations, such as weak-coupling and Markovian \cite{Breuer2002}, which are generally not satisfied in quantum biology or not justified for complex many-body quantum systems, where even the division between the open system and its environment may be somewhat arbitrary. Because of these considerations, effective models combining quantum and classical aspects of noise \cite{Rossi2017} while retaining strong physical insight and flexibility in the noise parameters are crucial for advancing our understanding on the role of quantumness in biological or chemical processes. 

In this paper, we introduce a new open quantum system approach that suitably combines quantum collision models and classical stochastic processes: the Stochastic Collision Model (SCM). Importantly, the model arises from a very intuitive physical description of decoherence as originating from stochastic collisions with a quantum environment composed of quantum ancillae. The freedom in the choice of the local stochastic processes allows us to model spatial and temporal heterogeneity in the noise. The dynamics of the system, after averaging over the stochastic realizations, describes both Markovian and non-Markovian behaviors, depending on the noise parameters, and does not make any assumption on the coupling strength between system and environment, therefore naturally going beyond the weak coupling limit. 

To demonstrate the usefulness, flexibility, and descriptive power of the SCM, we use it in the study of noisy transport in quantum networks, considering two paradigmatic examples: the fully connected graph and the Fenna-Matthews-Olson (FMO) complex \cite{FENNA1977751, Fenna1975}. In the first case, we bring to light the interplay between spatial/temporal noise heterogeneity and the presence of system-environment entanglement in the efficiency of transport. In the second case, we also focus on the features of noise optimality. We discover that optimality constraints unveil the existence of two classes of nodes. Specifically, transport efficiency is optimized when certain nodes are subjected to strong noise, while the others to very low levels. Remarkably, these two communities are consistent, to a high degree, with the recent discovery of two classes of site-dependent fluctuations in the FMO complex \cite{Saito2019}.

\section{The model}

The SCM is inspired by collision models, in which the system undergoes a series of unitary interactions with environmental ancillary qubits, resulting in an open system dynamics \cite{Scarani2002,Ziman2002,Giovannetti2012,Ciccarello2013,Ciccarello2017}. In the SCM, the ancillae collide with the individual parties conforming the system according to a stochastic process, that is, collisions occur at non-deterministic times. While this may a priori seem to be a minor difference with respect to deterministic-time collisions, it has important physical implications; in a previous publication~\cite{GarciaPerez2020}, we introduced a similar model for a single-qubit system, and we showed that the randomness in the collision times can result in the decoherence of the system qubit even in the absence of system-environment entanglement. While the model can naturally accommodate scenarios typically considered in the literature of collision models, such as allowing the ancillae to be initially correlated or to collide multiple times with the system, we will consider only the situation in which they are all initially in a product state and collide with the system only once. Therefore, we will mainly focus on the study of the effect of the collision dynamics on the system's behavior.

The SCM is a general noise model for systems composed of $N$ qubits driven by some Hamiltonian $H$, in which the effect of collisions at random times is added to the free evolution of the system. More precisely, if two consecutive collisions take place at times $t_1$ and $t_2$, the system evolves from $t_1$ to $t_2$ according to $\rho (t_2) = U \rho (t_1) U^{\dagger}$ with $U = e^{-i (t_2 - t_1) H }$. The collision between an ancilla --- initially in the ground state $| 0 \rangle$ --- and a system qubit $m$ is modeled through the unitary dynamics generated by the local interaction Hamiltonian $H_{\mathrm{anc},m}^{(I)} = (\eta / 2) \sigma_{\mathrm{anc}}^x \otimes \sigma_m^z$ during a short period of time $\tau$, after which the ancilla drifts away and never interacts with the system again. We assume the interaction time $\tau$ to be much shorter than any relevant time scale in the free system dynamics, so that collisions can be regarded as instantaneous processes resulting in the application of unitary transformations $U_m = e^{- i (\theta / 2)  \sigma_{\mathrm{anc}}^x \otimes \sigma_m^z}$, with $\theta =  \tau \eta$ being the interaction strength. This parameter regulates the entanglement between the system and the ancilla. For instance, for $\theta = (2 k +1)\pi, \, k \in \mathbb{Z}$, the interaction is non-entangling, while it can be maximally entangling for $\theta = (2 k +1)\pi / 2, \, k \in \mathbb{Z}$. In any case, the fact that the ancilla does not collide again with the system deems its degrees of freedom irrelevant after the collision event, so it can be safely ``traced out''. The effect of the composition of $U_m$ and the consequent partial trace results in a single quantum channel $\Phi_m \left[ \rho(t) \right] = K_m \rho(t) K_m^{\dagger} + K_m^{\dagger} \rho(t) K_m$, with $K_m = e^{-i (\theta / 2) \sigma_m^z} / \sqrt{2}$. While the system-ancilla interaction outlined here has been designed to account for dephasing, other interactions can result in dissipation through the same mechanism as well.

\begin{figure*}
    \centering
    \includegraphics[width=.8\textwidth]{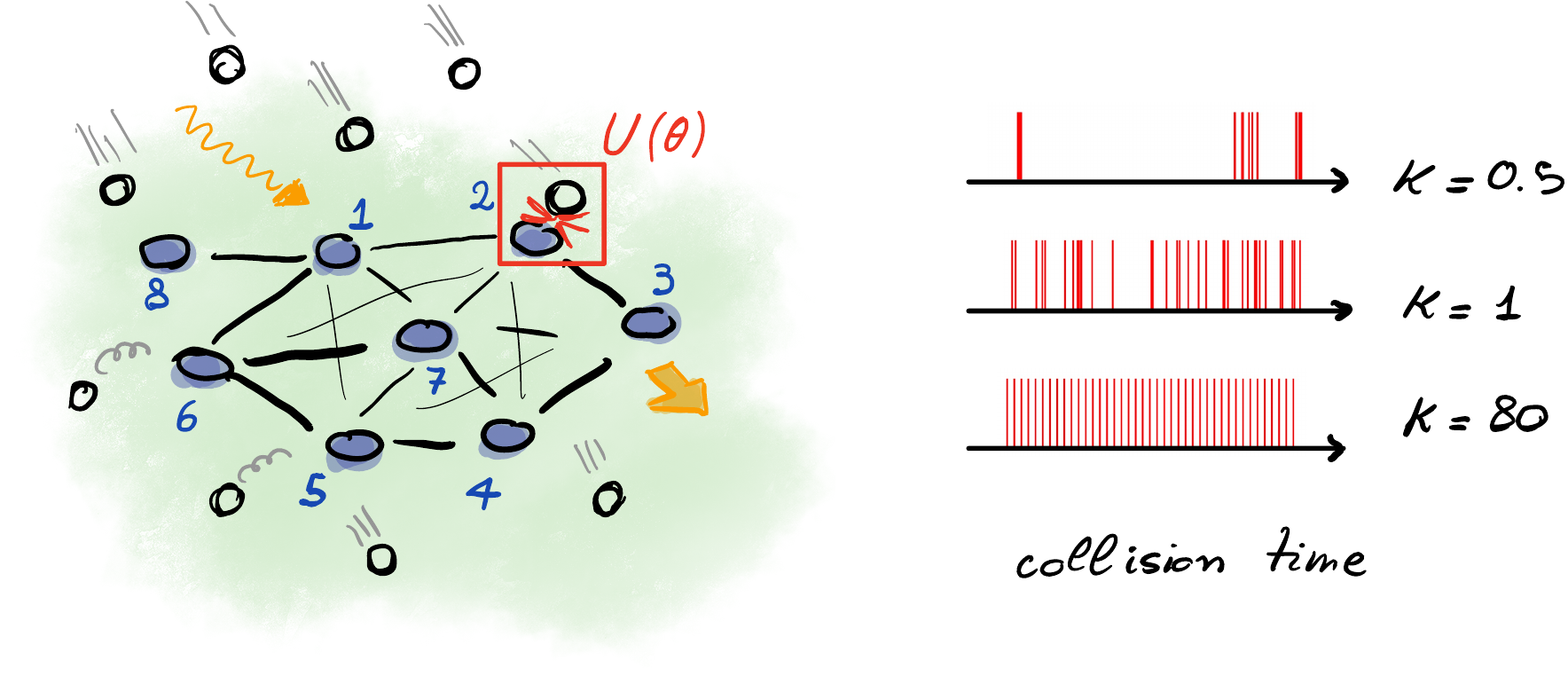}
    \caption{\textbf{Pictorial representation of the Stochastic Collision Model.} The sketch illustrates environmental ancillae colliding with the eight sites of the FMO complex. Each ancilla collides only once with one node of the FMO complex, while the time between two consecutive collisions follows the Weibull Renewal Process. On the right side we show how changing the value of $k \equiv k_i$ --- which for simplicity we assume here to be independent on the node $i$ --- we can describe the cases of heterogeneous ($k<1$),  Poisson ($k=1$),  and regular ($k\gg1$) collision time interval distribution.  }
    \label{fig:sketch}
\end{figure*} 

The collision time dynamics can in principle be any stochastic process of our choice. In this paper, we particularize to the case in which the collisions on all the system qubits occur independently following a Weibull Renewal Process (WRP)~\cite{Yannaros1994}. For every qubit $i$, the probability density for the interval between two consecutive collisions on $i$, $t_i$ is then given by a Weibull distribution,
\begin{equation}\label{eq:weibull_distribution}
p(t_i)=\frac{k_i}{\lambda_i}\bigg(\frac{t_i}{\lambda_i}\bigg)^{k_i-1}e^{-(t_i/\lambda_i)^{k_i}},
\end{equation}
in which $k_i$ and $\lambda_i$ are the shape and scale parameters, respectively. The main motivation for this choice is that it allows us to control the intensity and the heterogeneity in the collision statistics locally, that is, on each qubit independently. In particular, for a fixed value of $k_i$, the mean intercollision time $\langle t_i \rangle = \lambda_i \Gamma(1+1/k_i)$ (where $\Gamma$ stands for the gamma function) is proportional to $\lambda_i$, so we can control the \textit{spatial heterogeneity} of the noise across the system simply by choosing different scale parameters for each qubit. At the same time, the shape parameter enables the control of the \textit{temporal heterogeneity} of the noise, since $\mathrm{Var}(t_i) = \lambda_i^2 \left[  \Gamma(1+2/k_i) -  \Gamma(1+1/k_i)^2 \right]$. For $k_i < 1$, the collision dynamics is heterogeneous, characterized by bursts of collisions separated by long intervals of inactivity. As $k_i$ increases above 1, collisions become increasingly regular, as shown in Fig.~\ref{fig:sketch}. For $k_i = 1$, they follow a Poisson point process with rate $1 / \lambda_i$. Therefore, in that case, the dynamics of the system, when averaged over stochastic realizations, is described by the master equation
\begin{equation}\label{eq:master_equation}
\frac{d\rho(t)}{dt} = -i [H, \rho(t)] + \sum_{i=1}^{N}\lambda_{i}^{-1} \left\lbrace \Phi_{i} \left[ \rho(t) \right] - \rho(t) \right\rbrace,
\end{equation}
which is in Gorini-Kossakowski-Sudarshan-Lindblad form \cite{Lindblad1976, doi:10.1063/1.522979}, with $K_i$ and $K_i^{\dagger}$ the Lindblad operators. The system therefore undergoes Markovian dynamics. For values of $k_i$ different from 1, the collision dynamics is not memoryless~\footnote{Memorylessness refers to the property that, at any time $t$, the probability distribution for the waiting time until the next collision on a given node is independent of the time elapsed since the last collision with it.}, so the dynamical evolution cannot straightforwardly be written in Lindblad form. In all cases, however, the dynamics can be efficiently simulated by sampling the individual time intervals between collisions from Eq.~\eqref{eq:weibull_distribution} (see Appendix~\ref{app:scm_algorithm} for a description of the algorithm to simulate the SCM). In addition, the model could accommodate more general classical non-Markovian processes that can be simulated using recent computational techniques~\cite{Boguna:2014}.

In this paper, we apply the SCM to the study of transport phenomena in the framework of Continuous Time Quantum Walks (CTQW). In CTQW, one typically considers $N$ qubits interacting through a Hamiltonian that preserves the number of excitations, such as
\begin{equation}\label{eq:ctqw_hamiltonian}
    H = \sum_{i=1}^{N} \omega_{i} \sigma^{+}_{i}\sigma^{-}_{i} + \sum_{i \neq j} g_{ij} (\sigma_{i}^{+}\sigma_{j}^{-}\,+\,\sigma_{j}^{+}\sigma_{i}^{-}),
\end{equation}
where $\omega_{i} \in \mathbb{R}$ and $\sigma_{i}^{\pm}$ are the site energies and ladder operators of qubit $i$, respectively, and $g_{ij} \in \mathbb{R}$ are the hopping strengths between nodes $i$ and $j$. The fact that the hopping strengths can be set independently for every pair of qubits brings about an interpretation of the Hamiltonian in terms of a network, in which qubits are associated with its nodes and hopping strengths with weighted connections among them. Hence, the dynamics of a single excitation through the network --- a quantum walker --- is reminiscent of classical random walks on graphs. Furthermore, like in the case of classical random walks, the structural properties of the underlying network generally have non-trivial effects on the propagation of the quantum walker \cite{Muelken2011}.

Among the many fields in which CTQW find applications, they are widely used in quantum biology to model the propagation of energy across light-harvesting complexes \cite{Engel2007,Caruso2009,Chin_2010,Hoyer2010,Rossi2017a,Benedetti2019}. In these complexes, excitations must travel from a specific \textit{initial node} $r$ towards a \textit{target node} $s$, where they are finally captured. This last part of the process can be modeled by adding a sink, an extra node attached to the target node from which excitations decay irreversibly. Mathematically, this can be achieved without de facto increasing the dimension of the system by phenomenologically including in the Hamiltonian the non-Hermitian term $ - i \gamma\sigma^{+}_{s}\sigma^{-}_{s}$ , with $\gamma$ the \textit{sink rate}. Indeed, this term results in a leak of probability, which models the transfer of probability to the sink (its population). In other words, the sink's population at any time $t$ is given by $1 - \mathrm{Tr} \left[\rho(t)\right]$, where $\rho(t)$ is the state of the system at that time. The transfer dynamics is slowed down due to quantum Zeno effect for $\gamma\gg g_{ij}$ \cite{doi:10.1142/S1230161214400071}, while in the opposite regime $\gamma\ll g_{ij}$ the dynamics is effectively unitary.

The sink further allows us to define a figure of merit to quantify the effectiveness of the excitation transport. We define the \textit{performance} $\varepsilon$ as the inverse of the time required for the population of the sink to reach a certain value (which we set to 0.95). It should be noted that the fact that the evolution is non-unitary even in the absence of noise does not prevent us from applying the SCM to this system in any way.

\section{Results}
\subsection{Fully connected graph}
We first apply our noise model to study the energy transport in the fully connected network, where each node is connected with all others with the same hopping strength, $g_{ij} = g$, and the site energies are homogeneous (which we set to zero without loss of generality). The fully connected network is an interesting case study, as it is a system where the energy transfer is strongly suppressed in the absence of noise. In a network with $N$ nodes, only $1/(N-1)$ of the population of an initially localized excitation is able to reach the sink, most of it being trapped in the initial node~\cite{doi:10.1063/1.3223548}. This is because a localized state in a fully connected network is mostly composed of energy eigenstates with no spatial overlap with the sink. Such states evolve according to a unitary dynamics and are hence effectively decoupled from the sink (see Appendix \ref{app:fully_conn_dynamics}).

\begin{figure*}[t!]
    \centering
    \includegraphics[width=\textwidth]{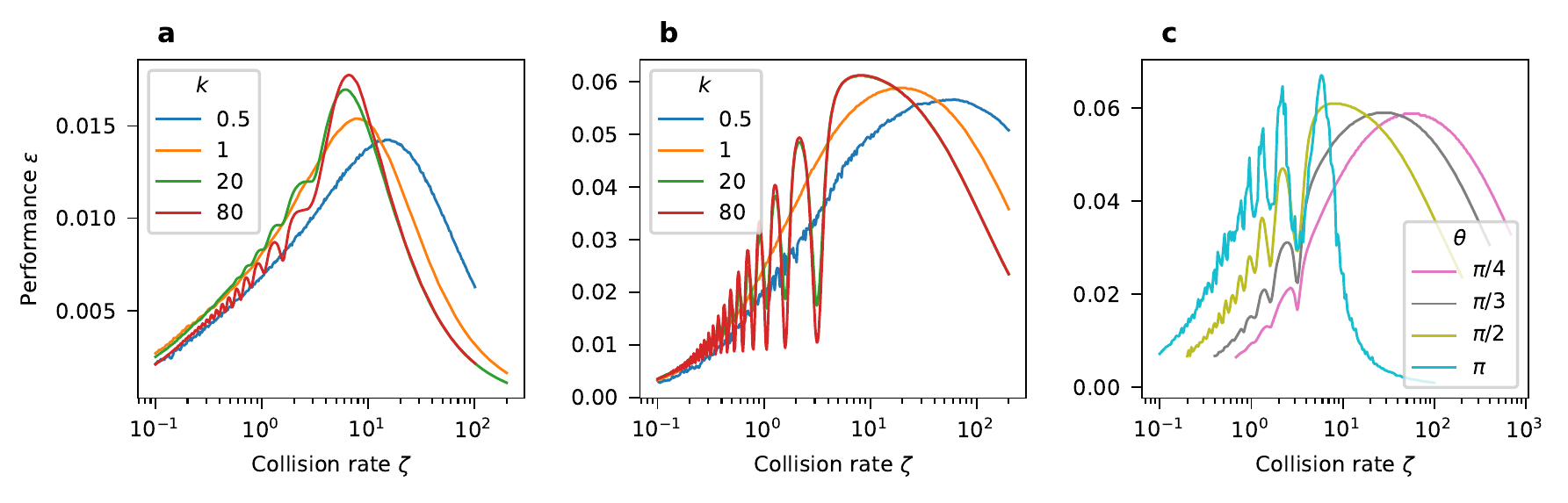}
    \caption{\textbf{Effect of noise heterogeneity and system-environment entanglement on transport performance.} In all cases, the curves show the performance $\varepsilon$ as a function of the inverse intercollision time $\zeta$ for a fully connected network with $N=20$ nodes. In \textbf{a}, ancillae collide with all nodes with equal shape and scale parameters. In \textbf{b}, collisions only take place with the initial node $r$. In both cases, $\theta = \pi/2$, and the legend indicates the shape parameter corresponding to each curve. In \textbf{c}, only node $r$ is affected by collisions, and $k_r = 10$. Each curve corresponds to a different interaction strength.}
\label{fig:fully_connected}
\end{figure*}

By adding dephasing noise, which maps states decoupled from the sink into mixtures of decoupled and coupled states, we can break the energy confinement and let the excitation flow through the network until it eventually reaches the sink. In Fig.~\ref{fig:fully_connected}, we show the performance $\varepsilon$ as a function of the \textit{collision rate} $\zeta \equiv 1 / \langle t_i \rangle$ for different spatial and temporal heterogeneity, as well as for different interaction strengths $\theta$. In particular, Fig.~\ref{fig:fully_connected}\textbf{a} corresponds to spatially homogeneous noise ($\lambda_i = \lambda, \, \forall i$), while Fig.~\ref{fig:fully_connected}\textbf{b} shows the results for maximally heterogeneous spatial noise, in which collisions only occur on the initial node $r$ (in this case, $\zeta = 1 / \langle t_r \rangle$). Figure \ref{fig:fully_connected}\textbf{c} shows the effect of the interaction strength for fixed spatial and temporal heterogeneity. Overall, the curves are characterized by a low performance for small collision rates, which are not effective in breaking the energy confinement, as well as for high rates, which slow down the excitation walk due to the Zeno effect. Hence, we observe an optimal collision rate $\zeta$, different in each case.

The comparison between Fig.~\ref{fig:fully_connected}\textbf{a} and \textbf{b} reveals the strong effect of spatial heterogeneity. While collisions on all the qubits are able to break the confinement, the process is much more efficient if the collisions are localized on the initial node $r$ only. Moreover, numerical investigation shows that localized noise on a single node that is not the initial one is instead a sub-optimal strategy. This behavior is a consequence of the fact that the initial node remains the most populated one throughout the dynamics (see Appendix \ref{app:fully_conn_dynamics}), which in turn implies that colliding with it is the most effective strategy to alter the coherences that give rise to the quantum interference-induced trapping. 

In addition to the spatial heterogeneity, the curves also portray the effect of temporal heterogeneity in the collision statistics. Both with spatially localized and homogeneous collisions, large values of $k_i$ give rise to a higher but narrower performance peak, while low $k_i$ yields lower and wider peaks, with good performances even for very high dephasing rates. Therefore, temporal heterogeneity makes the dynamics more resilient to the Zeno effect. Both homogeneity and heterogeneity in the collision times can thus be regarded as resources, depending of the circumstances.

Another evident and interesting phenomenon is that, in the case of large values of $k_i$, the performance exhibits sudden drops for some specific collision rates. The periodicity of the corresponding collision rates suggests that they are related to some characteristic time of the network, and their origin can indeed be easily understood by looking at the unitary dynamics of the system (i.e., in the absence of noise and sink). By introducing the single-excitation localized states $\left\lbrace \ket{i} \equiv \sigma_i^{+} \ket{0}^{\otimes n} \right\rbrace$, which form a basis of the single-excitation subspace (see Appendix~\ref{app:a}), the corresponding subspace Hamiltonian $H^{(1)}$ (that is, with matrix elements $H_{ij}^{(1)} =  \bra{i} H \ket{j}$) can be written as $H^{(1)} = g \left( N \ket{\phi} \bra{\phi} - \mathbb{1} \right) = g \left[ (N-1) \ket{\phi} \bra{\phi} - P_{\perp} \right]$, where $\ket{\phi} = \left( \sum_{i=1}^{N} \ket{i} \right) / \sqrt{N}$ and $P_{\perp} = \mathbb{1} - \ket{\phi} \bra{\phi}$ is the projector on the subspace orthogonal to $\ket{\phi}$. Consequently, the time evolution operator reads 
$e^{- i t H^{(1)}} = e^{igt} \left( e^{-igNt} \ket{\phi} \bra{\phi} + P_{\perp} \right)$, which equals identity (except for an irrelevant phase factor) at times
\begin{equation}\label{eq:drop_times}
	t=\frac{2 \pi m}{gN},\ \ \ m\in \mathbb{Z}.
\end{equation}
Therefore, in the limit of perfectly periodic collisions matching such periods, the ancillae always collide with a localized state --- the initial one --- with no effect, and the excitation remains trapped. In Fig.~\ref{fig:fully_connected}\textbf{b}, however, the performance does not drop to zero. The reason is that, while large values of $k_r$ result in nearly periodic collisions, they are ultimately random. In other words, even when the collision rate matches a characteristic time of the network, the actual collision times are noisy. Moreover, the width of the drops shows that a slightly detuned collision rate can still lead to a significant loss of performance, which suggests some robustness in this phenomenon. It should be mentioned that, when one of the nodes is attached to a sink, the state is not fully localized at times $t$ given by Eq.~\eqref{eq:drop_times}. Yet, at those times, the localization is highest, making the collisions as little effective as possible.

Finally, we also address the role of the interaction strength $\theta$. In Fig.~\ref{fig:fully_connected}\textbf{c}, we show the performance vs.~collision rate curves with fixed spatial and temporal heterogeneity (collisions occur only with the initial node $r$ with $k_r = 10$) for different values of $\theta$. We can appreciate that, as the interaction strength increases, the curves are shifted towards lower collision rates and, at the same time, the optimality peak narrows. Interestingly, the highest performance is achieved for non-entangling qubit-ancilla interactions, which do not cause decoherence in any single realization of the noise dynamics. We can also see the periodic performance drops in all cases, the positions of which do not depend on $\theta$ whatsoever. This is consistent with our explanation of the phenomenon in terms of the periodicity of the free network dynamics, according to which we expect to observe it as long as the collisions take place when the population is localized on $r$ and the interaction with such state has no effect (which is the case for any  value of $\theta$).

\subsection{FMO complex}

\begin{figure*}
    \centering
	\includegraphics[width=0.6\columnwidth]{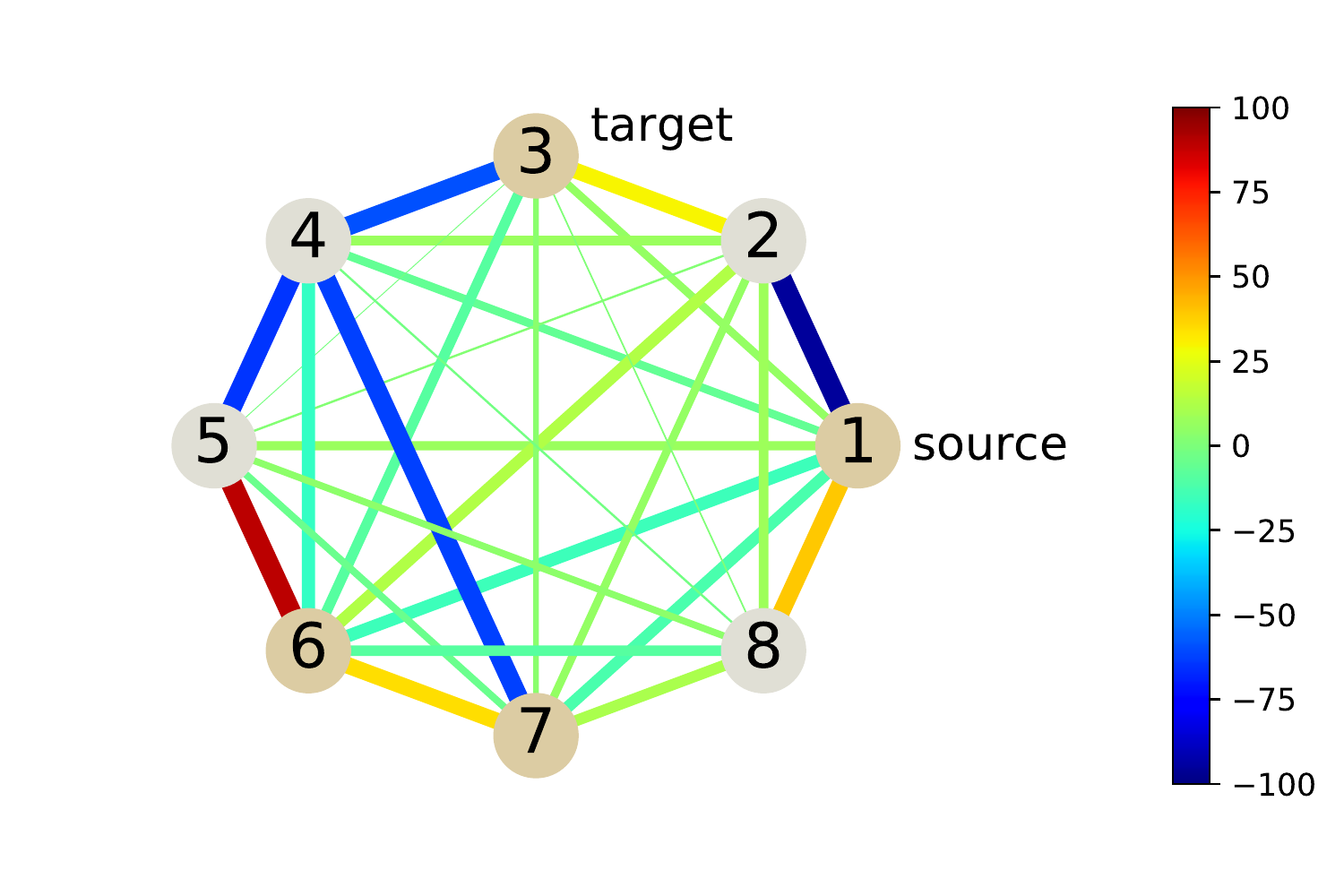}
    \caption{\textbf{Network representation of the FMO complex.} Nodes in the FMO are represented as circles, whereas the lines connecting them represent the hopping strengths. Every link's width is proportional to the logarithm of the corresponding matrix element, $\log \vert H_{ij}^{(1)} \vert$, while the color indicates the value according to the bar on the right. The colors of the nodes highlight their optimal noise level from Fig.~\ref{fig:fmo_optimal}: high (pale grey) or low (dark yellow).}
\label{fig:fmo_network}
\end{figure*}

We now turn our attention to the Fenna-Matthews-Olson (FMO) complex, a real networked system widely studied in the quantum biology literature. This complex appears in green sulfur bacteria and assists energy migration from a chlorosome super-complex to the reaction centre \cite{Jang2018,Cao2020}. It can be modeled as an 8-node fully connected network with non-homogeneous hopping strengths and non-homogeneous site energies. Using the values tabulated in Ref.~\cite{Jang2018}, we can write the matrix representation of its single-excitation subspace Hamiltonian in the computational basis (see Appendix \ref{app:a}) as
\begin{equation}
 H^{(1)} =
 	\left(\begin{matrix}
 	200 & -94.8& 5.5 & -5.9 & 7.1  & -15.1 & -12.2 & 39.5\\ 
 	-94.8& 230 & 29.8 & 7.6 & 1.6 & 13.1 & 5.7 & 7.9\\
 	5.5 & 29.8 & 0 & -58.9& -1.2 & -9.3 & 3.4 & 1.4\\
 	-5.9 & 7.6 & -58.9& 180 & -64.1& -17.4& -62.3 & -1.6\\
 	7.1  & 1.6 & -1.2 & -64.1& 405 & 89.5 & -4.6 & 4.4\\
 	-15.1& 13.1 & -9.3 & -17.4& 89.5 & 320 & 35.1 & -9.1\\
 	-12.2& 5.7 & 3.4 & -62.3& -4.6 & 35.1 & 270 & 11.1\\
 	39.5 & 7.9 & 1.4 & -1.6 & 4.4 & -9.1 & 11.1 & 505\\   
 	\end{matrix}\right).
\end{equation}
The site energies have been shifted so that the 3rd node, to which the sink is connected, has zero energy. A network representation of the complex is depicted in Fig.~\ref{fig:fmo_network}. In many previous works \cite{doi:10.1063/1.3223548, Chin_2010, Hoyer2010, PhysRevA.81.062346, Plenio_2008} the FMO is modeled as a 7-node network, as the eighth node has recently been discovered~\cite{SchmidtamBusch2011, Tronrud2009,Jang2018}. In any case, the two networks have a highly similar topology and behavior, and all the results presented here are valid for both of them. It is worth clarifying that the FMO is a large biomolecule with a complex internal structure that interacts with a warm environment. Its purpose is to transfer excitations as fast as possible to the reaction center before it is lost to the environment due to dissipation. Modeling the FMO as an 8-node network, the energy transfer as a CTQW within the network, and the environmental interaction as pure dephasing --- ignoring dissipation --- all corresponds to an effective model. While such models are capable of capturing the core behavior of these systems~\cite{Moix2011, doi:10.1063/1.4762839, Wilkins2015}, they are not adequate for accurate numerical computations.

\begin{figure*} [t]
    \centering
    \includegraphics{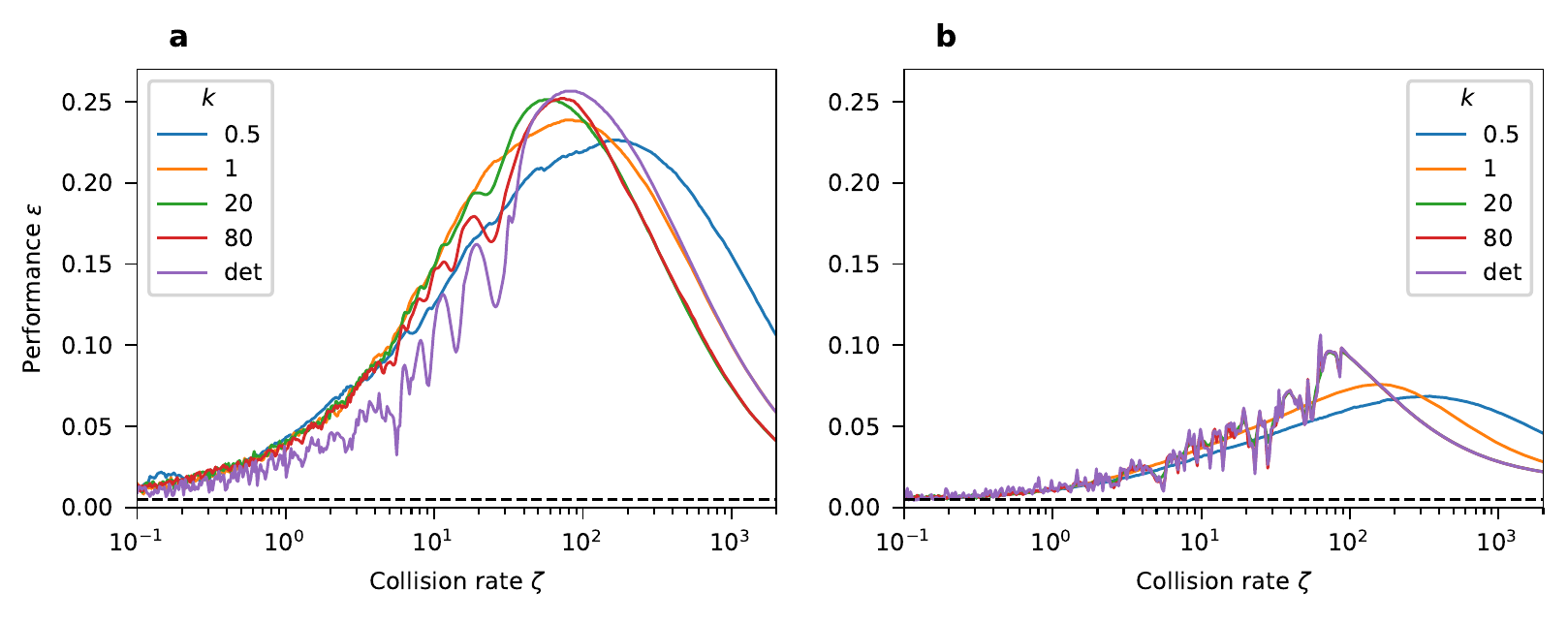}
    \caption{\textbf{Effect of noise heterogeneity on transport performance in the FMO complex.} Performance $\varepsilon$ as a function of the inverse intercollision time $\zeta$ in the case of \textbf{a} spatially homogeneous noise and \textbf{b} maximally heterogeneous noise (collisions on the initial node $r=1$ only). In both cases, $\theta = \pi/2$, and the legend indicates the shape parameter corresponding to each curve. The black dashed line shows the performance of FMO in the noiseless case.}
\label{fig:fmo_heterogeneity}
\end{figure*}

We apply the SCM to the FMO with different spatial and temporal heterogeneity, as we did with the fully connected network. By comparing the two extremes of spatial homogeneity (collisions on all nodes) and heterogeneity (collisions only on the initial node $r=1$) in Fig.~\ref{fig:fmo_heterogeneity}, we see that, in this system, localizing the collisions on the source node no longer is the most efficient strategy to drive the excitation to the sink. The FMO is a disordered system that does not experience strong population trapping in the initial node, so there is a priori no reason to expect localized dephasing to result in a better performance in this case. On the other hand, temporal heterogeneity has the same effect on the FMO as on the fully connected network: it leads to lower performance at optimal noise rates while resulting in increased resilience against the Zeno effect. 

Surprisingly, we also observe drops in the performance of the process for some collision rates in this system. In the fully connected network, this was associated with the presence of characteristic times in the dynamics, which could be identified given the simplicity of the system. Finding the characteristic times of the FMO from the free dynamics would be far from trivial, but these performance drops signal their existence in such a disordered system, and raises questions regarding the generality of the phenomenon. 
Given that the SCM is able to reveal this periodicity in the FMO dynamics, we now consider whether the versatility of the model can be further exploited to identify other properties of this real system. We have so far studied spatial heterogeneity by simulating the dynamics in two extreme cases, and the temporal one by jointly modifying the shape parameters of all nodes with non-zero collision rate. We now address the question of what the local scale and shape parameters maximizing the performance in the FMO are. To find the optimal values of $\lbrace \lambda_i, k_i \rbrace$, we use a genetic optimization algorithm \cite{Goldberg1989}. Essentially, the optimization is performed by considering a pool of candidate parameter assignments (the first generation), from which the best-performing ones (fittest) are selected and combined to obtain the next generation. This process is applied iteratively, yielding high-performing generations after several steps. A more detailed explanation of the algorithm can be found in Appendix~\ref{app:genetic_algorithm}.

\begin{figure} [t!]
    \centering
	\includegraphics{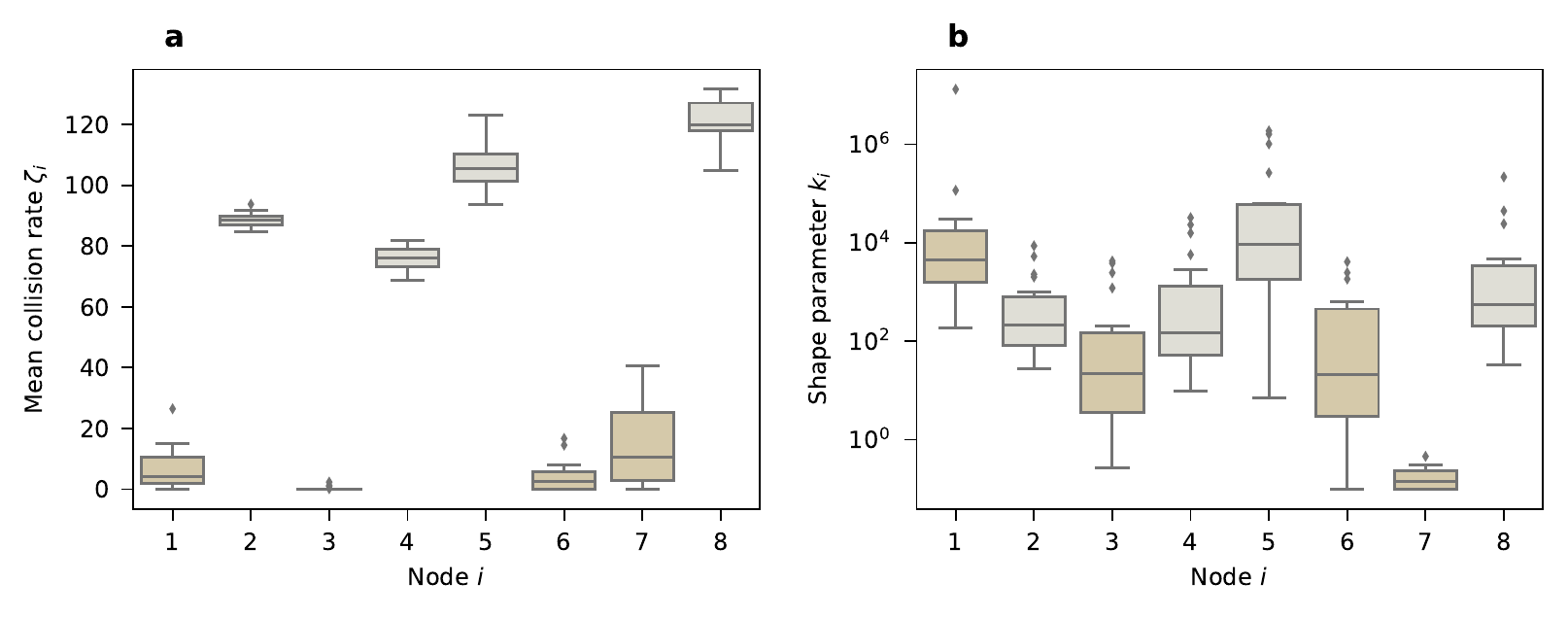}
	\caption{\textbf{Optimal noise for the FMO complex.} Distribution of \textbf{a} the mean collision rate $\zeta_i$, and \textbf{b} the shape parameters $k_i$ of node $i$ in the last generation pool. The boxes show the first and third quartile, the middle bar being the median. The whiskers show an inter-quartile range of 1.5. Panel \textbf{a} shows that there are two classes of nodes (highlighted with two different colors): one with weak noise (small $\zeta_i$) (nodes 1, 3, 6 and 7), and one with strong noise (large $\zeta_i$). The optimal shape parameters $k_i$ tend instead to be quite large, meaning that the collisions should be essentially deterministic.}
\label{fig:fmo_optimal}
\end{figure}

The results of the optimization are displayed in Fig.~\ref{fig:fmo_optimal}, where we show the distribution of the optimal mean collision rate $\zeta_i$, Fig.~\ref{fig:fmo_optimal}\textbf{a}, and of the shape parameters $k_i$, Fig.~\ref{fig:fmo_optimal}\textbf{b}, for each node. Interestingly, we notice two clearly distinct behaviors, with nodes being subjected to either very strong noise (very high collision rate) or virtually no noise at all. Moreover, Fig.~\ref{fig:fmo_optimal}\textbf{b} reveals that the noise is essentially deterministic. This result allows us to classify the nodes in the FMO into two classes according to their optimal SCM noise levels, and suggests a methodology for transport-based community detection \cite{Faccin2014}. It should also be stressed that this node classification is by no means evident from the structure of the network (see Fig.~\ref{fig:fmo_network}). 

\begin{figure} [t!]
    \centering
    \includegraphics{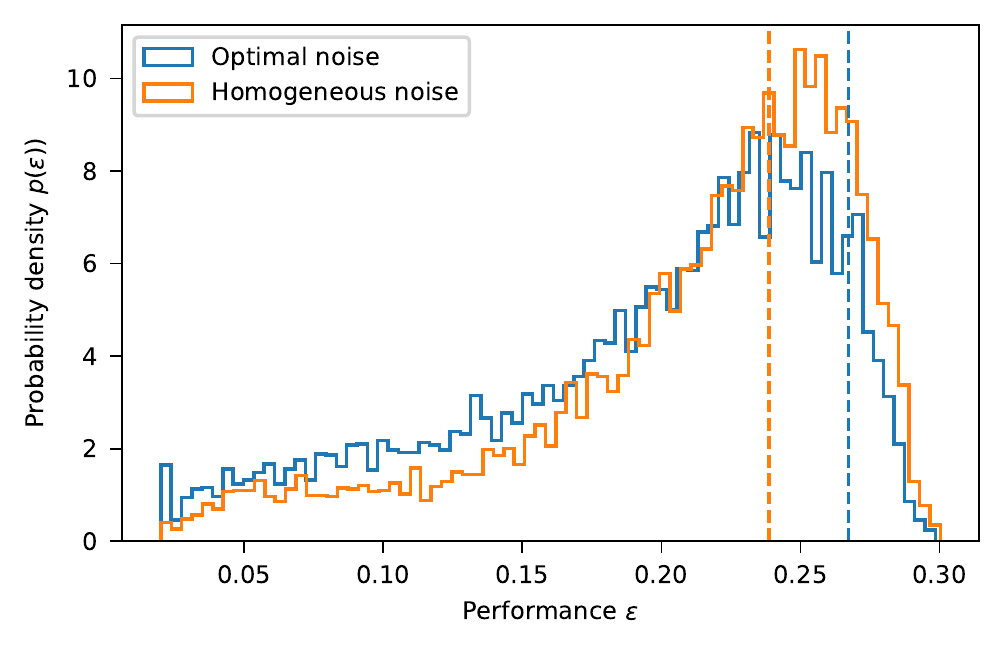}
	\caption{\textbf{Performance of the FMO upon link reshuffling.} The histograms shows the distribution of transport performance $\varepsilon$ for $10^4$ networks obtained by randomly reshuffling the links of the FMO. The orange histogram corresponds to spatially homogeneous and temporal Poissonian noise whereas, for the blue one, nodes were subjected to the optimal noise of Fig.~\ref{fig:fmo_optimal}. The vertical dashed lines indicate the performance of the actual FMO complex under the corresponding noise. In the case of homogeneous noise, $40\%$ of the reshuffled graphs are more efficient. Instead, for optimal noise, this is less common ($9 \%$).}
	\label{fig:fmo_reshuffled}
\end{figure}

To conclude our SCM-based analysis of the FMO complex, we focus on the structure of the Hamiltonian. Given that it is a biological system, one would expect Eq.~\eqref{eq:ctqw_hamiltonian} to be the result of an evolutionary process driving the system towards an optimal performance in noisy environments, such that modifying the hopping strengths would lead to a lower performance. However, this does not seem to be the case. In Fig.~\ref{fig:fmo_reshuffled}, we show the distribution of transport performance of networks obtained by randomly reshuffling (permuting) the hopping strengths --- so that all the networks have exactly the same weight distribution ---, along with the performance of the FMO. When subjected to homogeneous noise, as many as $40\%$ of these random networks outperform the original Hamiltonian. Similar results (not shown) can be obtained by simply sampling the hopping strengths independently from a Gaussian distribution with the same mean and variance. This phenomenon depends on the noise model as well. As shown in the same figure, the fraction of networks outperforming the FMO is significantly reduced when subjected to the optimal noise from Fig.~\ref{fig:fmo_optimal}. These rather surprising results do not necessarily contradict the assumption of the optimality of the FMO for excitation transfer. Although it seems relatively easy to outperform the actual FMO when allowed to choose the hopping rates freely, the actual physical system is subjected to constraints with which these randomized networks may be incompatible.

\section{Conclusions}
We have introduced the Stochastic Collision Model, a versatile noise model for general $N$-qubit systems in which ancillae collide with the system qubits at random times. We have moreover shown that by letting the collision-time dynamics on each qubit be driven by a Weibull Renewal Process, the model is conferred with the capability to regulate the spatial and temporal heterogeneity of the noise.

We have provided a thorough analysis of the effects of the noise heterogeneity on excitation transport in quantum networks both for the fully connected graph and the FMO light-harvesting complex. In the case of the fully connected network, the SCM reveals that, in the time-homogeneous regime, the transport performance is very sensitive to the periodicity of the system, a result for which we provide an explanation in terms of an analytical treatment of the noiseless dynamics. Remarkably, we observe a similar phenomenon in the FMO network, although with a significant difference: as opposed to the fully connected network case, this phenomenon is more evident, and the overall performance is higher, when the noise is homogeneous in space. Nevertheless, the sudden changes in the performance with respect to small variations in the collision rate suggest the existence of relevant periodic coherence-induced excitation-trapping events in the transport dynamics of this real system. While this stands as a conjecture at this point, this is a relevant aspect that requires further research.

We have also exploited the versatility of our model to explore other non-trivial effects of the FMO structure on its dynamical properties. By optimizing the transport performance over the noise parameters, we find that the best environmental conditions for this biological complex is given by high-rate nearly periodic collisions on some of its qubits, and no noise whatsoever on the rest. This surprising result does not seem to be expected from the structure of the underlying graph. Finally, we have addressed the question of whether the FMO structure is dynamically optimal, and we have found the unexpected result that a mere randomization of the hopping strengths often results in a more efficient system under homogeneous noise. However, when considering the optimal noise, the fraction of networks outperforming the FMO is drastically reduced.

Overall, our results showcase the potential of the SCM for several purposes. On the one hand, considering an interaction with the environment in terms of collisions with the particles conforming it is a sensible assumption in many relevant situations. Needless to say, in these scenarios, collision events occur randomly in time. In its more general form, the SCM can accommodate a wide class of stochastic dynamics, for which efficient simulation techniques exist. On the other hand, the model can be useful for the analysis of general systems in which it may not seem physically adequate, as one can use parameter-optimization techniques like the ones employed here to identify unknown dynamical phenomena in complex quantum systems.

\begin{acknowledgments}
G.G-P., M.R., and S.M. acknowledge financial support from the Academy of Finland via the Centre of Excellence program (Project no.~312058). D.A.C and G.M.P.~acknowledge PRIN project 2017SRN-BRK QUSHIP funded by MIUR. G.G.-P.~acknowledges support from the emmy.network foundation under the aegis of the Fondation de Luxembourg.
\end{acknowledgments}

\appendix
\section{Single-excitation subspace}
\label{app:a}

The Hamiltonian of the system, Eq.~\eqref{eq:ctqw_hamiltonian}, preserves the number of excitations. This means that, for any state with $k$ excitations $\ket{\psi_k} = \sigma_{i_1}^{+} \cdots \sigma_{i_k}^{+} \ket{0}^{\otimes N}$, the vector $H \ket{\psi_k}$ is orthogonal to any state with a number of excitations different from $k$. As a result, the Hamiltonian can be written as a direct sum of operators, each living in a $k$-excitation subspace, that is, $H = \bigoplus_{k = 0}^{N} H^{(k)}$. This significantly reduces the computational complexity of the problem, as we do not need to consider the full $2^{N}$-dimensional Hilbert space to study the dynamics of an initially localized single excitation. Instead, we can restrict our attention to the $N$-dimensional single-excitation subspace and its corresponding Hamiltonian $H^{(1)}$. It is convenient to introduce the basis of single-excitation localized states $\left\lbrace \ket{i} \equiv \sigma_i^{+} \ket{0}^{\otimes N} \right\rbrace$, in which the matrix representation of $H^{(1)}$ has elements $H_{ij}^{(1)} = \bra{i} H \ket{j} = g_{ij} (1 - \delta_{ij} ) + w_i \delta_{ij}$, which coincides with the adjacency matrix of the graph. In case the sink is considered, an extra term $-i \gamma \delta_{ij} \delta_{is}$ must be included. With the introduction of the sink, the Hamiltonian becomes non-Hermitian and does not preserve the population. This however does not prevent the use of the single excitation subspace. The loss of probability accounts for the population of the vacuum state, which is completely impervious to both the free evolution and the collisions, so it can be safely ignored.

\section{Noiseless dynamics in the fully connected network}
\label{app:fully_conn_dynamics}

In this section, we briefly outline some results regarding the dynamics of a single excitation in the fully connected network in the absence of collisions.

Using the expression for the time-evolution operator in the absence of sink and collisions presented in the main text, $e^{- i t H^{(1)}} = e^{igt} \left( e^{-igNt} \ket{\phi} \bra{\phi} + P_{\perp} \right) = e^{igt} \left[ (e^{-igNt} - 1) \ket{\phi} \bra{\phi} + \mathbb{1} \right]$, we see that the state of an excitation initially localized at node $r$ at time $t$ is $e^{- i t H^{(1)}} \ket{r} = e^{igt} \left[ (e^{-igNt} - 1) \frac{1}{\sqrt{N}}\ket{\phi} + \ket{r} \right]$. The population on some node $k$ is thus
\begin{equation}
    \left\vert \bra{k} e^{- i t H^{(1)}} \ket{r} \right\vert^2 = \delta_{rk} \left[ 1 + \frac{2}{N} \left( \cos(gNt) - 1 \right) \right] + \frac{2}{N^2} \left[ 1 - \cos(gNt) \right].
\end{equation}
Since the term multiplying the Kronecker delta is positive for $N>4$, we see that, for large networks, the initial node has the largest population throughout the dynamics.

Let us now turn our attention to the interference-induced coherence trapping, which results in a strong suppression of the energy transfer to the sink in this graph. The same argument that we present here can be found in \cite{Caruso2009}. As before, we consider an excitation initially localized on node $r$ and the sink to be attached to node $s$. If we ignore the sink for a moment, we can see that $\ket{\varphi_{i}}=\ket{r}-\ket{i}, \forall i \notin \lbrace r, s \rbrace$, along with $\ket{\phi}$, form a complete set of eigenstates of the Hamiltonian. Moreover, it is easy to show that 
\begin{equation}\label{eq:initial_node}
\ket{r} = \frac{1}{N-1}\left(\sum\limits_{i \notin \lbrace r, s \rbrace} \ket{\varphi_{i}} + \sqrt{N} \ket{\phi} - \ket{s} \right).
\end{equation}
The vector $\sum_{i \notin \lbrace r, s \rbrace} \ket{\varphi_{i}}$ is an eigenstate of the Hamiltonian and has no overlap with the target node state $\ket{s}$. Hence, its population is protected from the sink, while the population on the other two terms is unprotected and can eventually flow to the sink. Given that $\braket{\varphi_{i}}{\varphi_{j}} = 1 + \delta_{ij}$, we can write $\sum_{i \notin \lbrace r, s \rbrace} \ket{\varphi_{i}} = \sqrt{(N-1)(N-2)} \ket{\Psi}$ with $\braket{\Psi} = 1$. Introducing this into Eq.~\eqref{eq:initial_node}, yields
\begin{equation}
    \ket{r} = \sqrt{\frac{N-2}{N-1}}\ket{\Psi}+\ket{\mathrm{unprotected}}.
\end{equation}
The unprotected population, that is, the maximum amount of population that can reach the sink, can be readily computed from the expression above as $\braket{\mathrm{unprotected}} = \braket{r} - (N-2)/(N-1) \braket{\Psi} = 1/(N-1)$ (where we have used $\braket{\Psi}{\mathrm{unprotected}}= 0$ since $\braket{\varphi_i}{\phi}= 0$ and $\braket{\varphi_i}{s}= 0$). This trapping of coherence is due to the peculiar eigenstate structure of the Hamiltonian, which is in turn a consequence of the high symmetry of the fully connected network.

\section{Simulating the SCM}
The dynamics of a system subjected to a SCM can be evaluated by averaging over all possible realizations. This may not be easy to do analytically, but can always be numerically simulated. 

The time interval between two collisions on the same node are sampled from its waiting time distribution, Eq.~\eqref{eq:weibull_distribution}, that is in principle different for each node. If we want to sample one possible realization of the network collision times it is sufficient to extract the collision times for each node and order them chronologically.
Whenever there is a collision with a node $m$, the system is subjected to the collision map $\Phi_m$, while it undergoes free evolution between collisions.
\label{app:scm_algorithm}

\section{Genetic algorithm}
The purpose of the genetic algorithm is to find the noise parameters that yield the highest performance for the given network. 
An initial population is created by sampling random values. The population is composed by a certain number of members, each consisting of the set of parameters (called chromosomes) that we wish to optimize. The performance of each member of the population is evaluated, and the best half is promoted to be the parents of the next generation. The values of the parents are mixed to create an offspring, and the offspring is further subjected to random mutations. The parents and the offspring are now the members of the second generation, whose performance is evaluated and the process begins anew. By including the parents in the future generation we ensure that high performing members are never lost (the offspring do not always outperform their parents). This way, through each generation, we improve the overall quality of the genetic pool.

In our case we used a pool of 40 individuals, 20 of which were selected to be the parents at each iteration. Each individual corresponds to the shape and scale noise parameters for each qubit of the networks, so the total number of chromosomes is 16 when optimizing for the FMO. Each offspring individual inherits 8 chromosomes from one parent and 8 from the other, and each parent mates twice with different partners. The mutation usually consists in adding randomly extracted values to the chromosomes, though we often found faster convergences by multiplying by random values instead, or using a hybrid model. The number of generations is not fixed, but rather the algorithm is stopped once the quality of the population reaches a plateau.
\label{app:genetic_algorithm}

\bibliography{Refer}
\end{document}